\documentstyle{article}
\begin{document}
\title{Dirty black holes: \\
       Entropy as a surface term}
\author{Matt Visser\cite{e-mail}\\
        Physics Department\\
	Washington University\\
	St. Louis\\
	Missouri 63130-4899}
\date{30 July 1993}
\maketitle
\begin{abstract}
It is by now clear that the naive rule for the entropy of a black
hole, $(\hbox{entropy}) = 1/4 \; (\hbox{area of event horizon})$,
is violated in many interesting cases.  Indeed, several authors
have recently conjectured that in general the entropy of a dirty
black hole might be given purely in terms of some surface integral
over the event horizon of that black hole.  A formal proof of this
conjecture, using Lorentzian signature techniques, has recently
been provided by Wald.  This note performs two functions.  Firstly,
by extending a previous analysis due to the present author [Physical
Review D48, ???? (15 July 1993)] it is possible to provide a rather
different proof of this result --- a proof based on Euclidean
signature techniques. The proof applies both to arbitrary static
[aspheric] black holes, and also to arbitrary stationary axisymmetric
black holes. The total entropy is
\[
S = {k A_H\over4 \ell_P^2}
  +  \int_H  {\cal S} \; \sqrt{ {}_2 g}\; d^2x.
\]
The integration runs over a spacelike cross-section of the event
horizon $H$. The surface entropy density, ${\cal S}$, is related
to the behaviour of the matter Lagrangian under time dilations.
Secondly, I shall consider the specific case of Einstein-Hilbert
gravity coupled to an effective Lagrangian that is an arbitrary
function of the Riemann tensor (though not of its derivatives). In
this case a more explicit result is obtained
\[
S = {k A_H\over4\ell_P^2}
  + 4\pi {k\over\hbar} \int_H \;
         {\partial{\cal L} \over \partial R_{\mu\nu\lambda\rho}} \;
	 g^\perp_{\mu\lambda}  g^\perp_{\nu\rho}\;
	\sqrt{ {}_2 g} d^2x .
\]
The symbol $g^\perp_{\mu\nu}$ denotes the projection onto the
two-dimensional subspace orthogonal to the event horizon.  Though
the derivation exhibited in this note proceeds via Euclidean
signature techniques the result can be checked against certain
special cases previously obtained by other techniques, {\sl e.g.}
$(Ricci)^n$ gravity, $R^n$ gravity, and Lovelock gravity.
\end{abstract}

\newpage
\section{INTRODUCTION}

The entropy versus area relationship for generic ``dirty''  black
holes has recently engendered considerable interest
\cite{Visser93,JM:93a,JM:talk,JM:93b,Wald:talk,Wald93}.  Generically
a dirty black hole~\cite{Visser92} is a black hole distorted by
either:  (1) various classical matter fields, (2) higher curvature
terms in the gravity Lagrangian [{\sl e.g.} $(Riemann)^n$], or (3)
infestation with some version of quantum hair.

The present note addresses two main points:

Firstly, it has recently been conjectured that the entropy of a
dirty black hole can {\sl always} be cast into the form of an
integral of some quantity over the event
horizon~\cite{JM:talk,Moss-Myers,Simon-Whiting}.  A formal proof
of this result, based on Lorentzian signature Lagrangian techniques,
has recently been announced~\cite{Wald:talk}. Details and applications
may be found in~\cite{Wald93,Wald-Iyer}. In this note I present an
alternative proof of this result. The present proof is obtained by
utilizing Euclidean space techniques in the manner of~\cite{Visser93},
and is ultimately an extension of the original Gibbons-Hawking
Euclidean signature technology~\cite{Gibbons-Hawking}.  The proof
applies both to arbitrary static [aspheric] black holes, and also
(with additional technical complications) to arbitrary stationary
axisymmetric black holes. The total entropy is
\begin{equation}
S = {k A_H\over4 \ell_P^2}
  +  \int_H  {\cal S} \; \sqrt{ {}_2 g}\; d^2x.
\end{equation}
The integration runs over a spacelike cross-section of the event
horizon $H$. The surface entropy density, ${\cal S}$, is related
(in a particular manner involving time dilations) to the surface
term arising in the integration by parts that connects the
stress-energy tensor with the variation of the action under a
variation of the spacetime metric.  For definiteness, the calculations
are carried out in four-dimensional spacetime, but the generalization
to arbitrary dimensionality is immediate.

Secondly, as a specific example, this note will focus on the case
of black holes in Einstein-Hilbert gravity coupled to an effective
Lagrangian that is any arbitrary function of the Riemann tensor
(though not of its derivatives). Interest in such a toy model is
justified by noting that whatever the underlying quantum theory of
gravity is, one would expect on general grounds that the low energy
theory should be describable by an effective Lagrangian  that
contains at least the class of terms indicated above. Applying the
general formalism developed in this note to this particular case
yields
\begin{equation}
S = {k A_H\over4\ell_P^2}
    + 4\pi {k\over\hbar} \int_H \;
         J^{\mu\nu\lambda\rho} \;
	 g^\perp_{\mu\lambda}  g^\perp_{\nu\rho}\;
	\sqrt{ {}_2 g} d^2x.
\end{equation}
The tensor $J^{\mu\nu\lambda\rho} \equiv \partial{\cal L} /
\partial R_{\mu\nu\lambda\rho}$ has the same symmetries at the
Riemann tensor. The symbol $g^\perp_{\mu\nu}$ denotes the projection
onto the two-dimensional subspace orthogonal to the event horizon.

It is instructive to check this formula against several special
cases that have been derived by rather different methods.  For
instance, Jacobson and Myers~\cite{JM:93a} have recently evaluated
the entropy for black holes in Lovelock gravity using Hamiltonian
methods. The present analysis reproduces their result with no
difficulty. More recently, Jacobson, Kang, and Myers~\cite{JM:talk,JM:93b}
have extended their analysis to the case where the Lagrangian is
an arbitrary function of the Ricci scalar.  The entropy for black
holes of this type was extracted by using a combination of field
redefinition and Hamiltonian techniques. [Consider the behaviour
of the black hole under conformal deformations.] Again, this result
can be shown to be a special case of the general formula given
above. Furthermore, Jacobson, Kang, and Myers~\cite{JM:talk,JM:93b}
have also considered the case where the Lagrangian is the
Einstein-Hilbert Lagrangian augmented by the square of the Ricci
tensor. The present techniques allow a simple extension of that
result to the case of an arbitrary function of the Ricci tensor.
The fact that different techniques give the same answer where they
overlap is encouraging.

\underbar{Notation:} Adopt units where $c\equiv 1$, but all other
quantities retain their usual dimensionalities, so that in particular
$G\equiv \ell_P/m_P \equiv \hbar/m_P^2 \equiv \ell_P^2/\hbar$. The
metric signature is $(+,+,+,+)$. The symbol $T$ will always denote
a temperature. The stress-energy tensor will be denoted by
$t^{\mu\nu}$, and its trace by $t$.

\section{GENERAL THEOREM}

\subsection{Reprise}

In a previous paper~\cite{Visser93}, I have derived a general formula
for the entropy of a dirty black hole in terms of: (1) the area of
the event horizon, $A_H$, (2) the energy density in the classical
fields surrounding the black hole, $\varrho$, (3) the Euclideanized
Lagrangian describing those fields, ${\cal L}$, (4) the Hawking
temperature, $T_H$, (5) the entropy density, $s$, associated with
the fluctuations [quantum hair, statistical hair], and finally (6)
the metric. The total entropy is:
\begin{equation}
S = {k A_H\over4 \ell_P^2}
  + {1\over T_H} \int_\Sigma \{ \varrho - {\cal L}  \} K^\mu d\Sigma_\mu
  + \int_\Sigma s V^\mu d\Sigma_\mu.
\end{equation}
This formula applies to all static black holes [not necessarily
spherically symmetric], and to stationary non-static [axisymmetric]
black holes.  $K^\mu$ is the timelike Killing vector. $V^\mu$ is
the four-velocity of a co-moving observer. For a static black hole,
this is just the four-velocity of a FIDO [fiducial observer]. For
a rotating black hole this is the four-velocity of a co-rotating
observer. $\Sigma$ denotes a constant time hypersurface. The first
term in this formula agrees with Bekenstein's original
suggestion~\cite{Bekenstein}, with the normalization constant fixed
by Hawking's calculation~\cite{Hawking-Radiation}.  For the time
being fluctuations are ignored, ($s=0$, no quantum hair, no
statistical mechanics effects).

The issue of interest is the evaluation of the term:
\begin{equation}
{1\over T_H} \int_\Sigma \{ \varrho - {\cal L} \} K^\mu d\Sigma_\mu
=  {k\over\hbar} \int_\Omega \{ \varrho - {\cal L}  \} \sqrt{g} d^4x.
\end{equation}
Here $\Omega$ denotes the entire Euclidean four-manifold. As is
usual in the Euclidean formulation the time direction is compact
with period $\hbar\beta = \hbar/(k T_H)$. The Hawking temperature,
$T_H$, is related to the surface gravity, $\kappa$,  by $k T_H =
\hbar\kappa/2\pi$.  By their very construction, Euclidean signature
techniques are capable of addressing only the equilibrium thermodynamics
of that class of black holes whose surface gravity is constant over
the event horizon. Consequently, the ``zeroth law'' of black hole
thermodynamics will be adopted {\sl by fiat}.

By judicious use of several integrations by parts this integral
will be transformed into a surface integral over the two-dimensional
event horizon. To show this one must first introduce some extra
technical machinery.

\subsection{Metric}

\subsubsection{Static geometry}

In the case of a static, possibly aspheric, black hole the Euclidean
signature metric can be cast into the form
\begin{equation}
g = + N^2 \; dt \otimes dt+ g_{ij} \; dx^i \otimes dx^j.
\end{equation}
The quantity $N$ is known as the lapse function. The event horizon
occurs at $N=0$. The timelike Killing Vector is given by $K \equiv
{\partial/\partial t}$. In coordinates $K^\mu = (1,0,0,0)$; $K_\mu
= (N^2,0,0,0)$. FIDOS [fiducial observers] follow integral curves
of the Killing vector, thus the four-velocity of a FIDO is $V \equiv
K/||K||$.  In coordinates $V^\mu = (1/N,0,0,0)$, $V_\mu = (N,0,0,0)$.

Consider the one-form $dt$. Note that $||dt|| = 1/N$. In coordinates
$(dt)_\mu = (1,0,0,0)$; $(dt)^\mu = (1/N^2,0,0,0)$.  Consequently
the one-form $dt$ and Killing vector $K$ are parallel, indeed $dt
= K/N^2$.

The four-acceleration of a FIDO is given by $a \equiv (V\cdot\nabla)V$.
In coordinates $a^\mu = V^\nu \nabla_\nu V^\mu = -(1/N) g^{\mu\nu}
\nabla_\nu N$. Define the unit normal to the constant lapse
hypersurface by $n^\mu$, then by construction $a^\mu = -||a||
n^\mu$.  Using the fact that the Killing vector is hypersurface
orthogonal, a brief computation shows
\begin{eqnarray}
\nabla_\mu K_\nu
&=& -{1\over N}
   \left( K_{\mu} \nabla_{\nu} N -  K_{\nu} \nabla_{\mu} N \right)
   \nonumber\\
&=& \left(  K_{\mu} a_{\nu} - a_{\mu} K_{\nu} \right)
   \nonumber\\
&=& \{ ||a|| N \}
   \left( n_{\mu} V_{\nu} -  n_{\nu} V_{\mu} \right)_.
\end{eqnarray}
Furthermore $\nabla_\mu V^\nu = V_\mu a^\nu = -||a|| V_\mu n^\nu$.
The surface gravity is defined by $\kappa = \lim_H \{ N ||a|| \}
= \lim_H \{ ||\nabla N|| \} = \lim_H \{ \partial N/\partial\eta
\}$.

\subsubsection{Stationary geometry}

In the case of a stationary non-static black hole the Euclidean
signature metric can be put into the form (see {\sl e.g.}~\cite{Membrane})
\begin{equation}
g = + N^2 \; dt \otimes dt
    + g_{ij} \; (dx^i - \beta^i dt)\otimes (dx^j - \beta^j dt).
\end{equation}
In this more complicated situation it is possible to distinguish
at least four interesting classes of fiducial observers --- STATORS,
ZEVOS, ZAMOS, and ROTORS.

As previously, $N$ is known as the lapse function. The timelike
Killing Vector is still $K \equiv {\partial/\partial t}$. In
coordinates $K^\mu = (1,0,0,0)$.  A STATOR [stationary observer at
rest] is one who follows the integral curves of the Killing vector.
$V_S ^\mu = K^\mu/||K||$.  Note that $||K||^2 = N^2 +
g_{ij}\beta^i \beta^j$. In Lorentzian signature the vanishing of
$||K||$ defines the ergosphere, a concept that has no analogue in
Euclidean signature. The notion of a STATOR will not prove particularly
useful in what follows.

Consider the one-form $dt$. Even in a stationary [as opposed to
static] geometry it is still true that $||dt|| = 1/N$.  In either
Lorentzian or Euclidean signature the vanishing of $N = 1/||dt||$
defines the event horizon.  This one-form may be used to introduce
the notion of a ``minimally dragged '' observer --- a ZEVO [zero
vorticity observer]. A ZEVO is an observer whose (covariant)
four-velocity is defined to be $V_Z  = dt/||dt|| = N dt$. The
appelation is justified by calculating the vorticity of such a
system of observers:  $\varpi = *(V \wedge d V) = 0$.

In coordinates $V_Z^\mu = (1; \beta^i)/N$. Define the relevant
four-acceleration to be $a_Z  \equiv (V\cdot\nabla)V$. A brief
computation shows that the ZEVOS inherit much but not all of the
structure of the FIDOS of a static geometry. For instance $a_Z^\mu
= -(1/N) (g^{\mu\nu} - V^\mu V^\nu ) \nabla_\nu N$. The projection
operator is needed because $(V \cdot \nabla N) = (\beta^i \partial_i
N) /N \neq 0$, unless further assumptions are made.  The surface
gravity is defined by $\kappa = \lim_H \{ N ||a||\}= \lim_H
\{ ||\nabla_\perp N|| \}$. Note that if the stationary geometry is
in fact static that the system of ZEVOS coincides with the system
of STATORS and one recovers the system of FIDOS.

It is believed that every black hole that is stationary but not
static must be axially symmetric. Physically, the reason for this
is that a rotating ({\sl i.e.} nonstatic) black hole induces tidal
dissipation in any system that is not axially symmetric. The final
equilibrium state should thus be either static or axially symmetric.
While some rigorous theorems along these lines can be proved for
Einstein-Hilbert gravity the situation regarding more general
theories is far from clear.

Nevertheless, if one adopts these physical arguments above to
justify specializing to axial symmetry the metric may be further
reduced to the form (see {\sl e.g.}~\cite{Membrane})
\begin{equation}
g = + N^2 \; dt \otimes dt
    + g_{\phi\phi} \; (d\phi - \omega dt)\otimes (d\phi - \omega dt)
    + g_{AB} \; dx^A \otimes dx^B.
\end{equation}
There are now two Killing vectors, the timelike Killing vector
$K\equiv \partial/\partial t$, and the axial Killing vector $\tilde
K\equiv\partial/\partial\phi$.  [$K^\mu = (1,0,0,0)$; $\tilde K^\mu
= (0,0,0,1)$.] Because of the axial symmetry it is now possible to
define the notion of angular momentum.  The notion of the ``minimally
dragged'' ZEVO system discussed above now
particularizes to the notion of the ZAMO [zero angular momentum
observer]. For a ZAMO: $[V_\omega]^\mu = (1,0,0,\omega)/N$.
This implies $V_\omega  \propto  (K+\omega \tilde K)$. Note that
$||K+\omega \tilde K|| = N$. Thus $V_\omega  \equiv dt/||dt||
= (K+\omega \tilde K)/||K+\omega \tilde K||$.  Rearranging yields
the useful result
\begin{equation}
dt = (K+\omega \tilde K)/N^2.
\end{equation}
Because $V\cdot\nabla N \propto (K+\omega \tilde K)\cdot \nabla N
= 0$, the formulae for locally measured acceleration and surface
gravity simplify from those appropriate to the ``minimally dragged''
ZEVOS, and one has results more closely related to those of the
static FIDOS. For instance, one recovers $a^\mu = V^\nu \nabla_\nu
V^\mu = -(1/N) g^{\mu\nu} \nabla_\nu N$, while for the surface
gravity $\kappa = \lim_H \{ N ||a|| \} = \lim_H \{ ||\nabla N||
\}$.

Next, define the angular velocity of the event horizon by $\Omega_H
\equiv \lim_H \omega$.  For Kerr and Kerr-Newman black holes it is
possible to show, as a mathematical theorem, that $\Omega_H$ is a
constant everywhere on the event horizon. Indeed, for Kerr and
Kerr-Newman black holes, as one approaches the horizon $\omega =
\Omega_H + O(N^2)$. For arbitrary theories with arbitrary stress-energy
tensors the truth or falsity of such results is far from clear. To
obtain such results would require, at a  minimum, the use of the
field equations together with some form of the energy conditions.
[This parallels the question of the constancy of the surface gravity
over the event horizon.] As with the question of the surface gravity,
Euclidean signature techniques cannot even be set up unless $\Omega_H$
is a constant.  Physically, this is due to the fact that Euclidean
signature techniques are intrinsically limited to the analysis of
equilibrium thermodynamics.  If $\Omega_H$ is not a constant then
the implied differential rotation leads to shearing and dissipation
so that the situation is decidedly not in equilibrium.  Consequently
the constancy of $\Omega_H$ will be adopted {\sl by fiat}. [The
assumed constancy of $\Omega_H$ is equivalent to assuming that the
horizon of a stationary  axisymmetric black hole is a Killing
horizon, {\sl cf}~\cite{Wald93}.]

Having done this, it is now possible to introduce a fourth class
of fiducial observers --- the ROTORS [co-rotating observers].
Consider the Killing vector $K_\Omega = K + \Omega_H \tilde K$. In
coordinates $K_\Omega^\mu = (1,0,0,\Omega_H)$. Consequently
$||K_\Omega||^2 = N^2 + g_{\phi\phi} (\Omega_H - \omega)^2$. Thus
$K_\Omega$ is that unique Killing vector that is null on the event
horizon. The co-rotating observers are defined by $V_\Omega \equiv
K_\Omega/||K_\Omega||$. Note that the ROTOR system of co-rotating
observers and the ZAMO system have the same limit as one approaches
the horizon. For convenience I shall sometimes write $N_\Omega$
for $||K_\Omega||$. Note that both $N$ and $N_\Omega$ vanish on
the event horizon.

In Lorentzian signature the system of co-rotating fiducial observers
breaks down at sufficiently large distances.  [$K_\Omega$ becomes
spacelike for $\Omega_H r \geq c$.] There is no analogue of this
behaviour in Euclidean signature, and it can be safely ignored.

One has $a_\Omega \equiv (V_\Omega\cdot\nabla)V_\Omega =
(K_\Omega\cdot\nabla) K_\Omega/ ||K_\Omega||^2 =  -\nabla N_\Omega /
N_\Omega$. The surface gravity is given by $\kappa = \lim_H \{
N _\Omega ||a_\Omega|| \} = \lim_H \{ ||\nabla(N_\Omega)||
\}$.  That this definition in terms of ROTORS coincides with the
definition in terms of ZAMOS is yet another manifestation of the
fact that these two systems tend to the same limit at the event
horizon.

The necessity for this extended discussion of fiducial observers
arises from the fact these distinctions are both useful and necessary
for the following discussion. For static black holes it suffices
to use the simple system of FIDOS. For rotating black holes it is
the ROTOR system of co-rotating observers that plays a primary
role, first in defining the entropy, and secondly in performing
the manipulations to be discussed below. The ZAMO system is also
used, but is of secondary importance. It is to be emphasised that
whatever stress-energy is surrounding the black hole it must, by
the assumed internal equilibrium, be co-rotating with the hole.
That is, the four-velocity of the ROTOR system of co-rotating
observers must be an eigenvector of the stress-energy tensor.

\subsection{Action}

Take the Euclidean action to be
\begin{equation}
I_{tot}(g) =  I_{EH}(g) + I_m(g).
\end{equation}
The Einstein-Hilbert action is
\begin{equation}
I_{EH}(g) =
- {1\over16\pi G} \int_\Omega R \sqrt{g} \; d^4x
- {1\over8\pi G} \int_{\partial\Omega} K \sqrt{{}_3 g} \; d^3x,
\end{equation}
and consists of: (1) the original Einstein--Hilbert Lagrangian, to
be integrated over the entire Euclidean manifold, and (2) the
Gibbons-Hawking surface term~\cite{Gibbons-Hawking}. Here ${}_3 g$
denotes the induced three-metric on the three-dimensional hypersurface
$\partial\Omega$, while $K$ denotes the trace of the second
fundamental form.  By the assumed asymptotic flatness of the black
hole spacetime this term is to be integrated only over the three
surface at spatial infinity~\cite{Gibbons-Hawking,Hawking:Centenary}.

For an arbitrary variation of the metric
\begin{equation}
\delta I_{EH}(g) =
 {1\over16\pi G} \int_\Omega G^{\mu\nu} \delta(g_{\mu\nu}) \sqrt{g} \; d^4x
 -  \int_{\partial\Omega} \Theta_{EH}(\delta g) \sqrt{{}_3 g} \; d^3x.
\end{equation}
The surface term, $\Theta_{EH}$ depends in a linear
fashion on $\delta g$ and its first derivative. For the augmented
Einstein-Hilbert action, it is a special case result that these surface
terms $\Theta_{EH}$ vanish provided that $\delta g$, though not
necessarily its normal derivative, vanishes on the boundary.

The ``matter'' action is of the form
\begin{equation}
I_m(g) =
 \int_\Omega  {\cal L} \sqrt{g} \; d^4x.
\end{equation}
Here ${\cal L}$ denotes the Euclideanized ``matter'' Lagrangian
(All higher order geometrical terms [{\sl e.g.}  $(Riemann)^n$]
are lumped into this ``matter'' Lagrangian.)

For an arbitrary variation of the metric
\begin{equation}
\delta I_{m}(g)
= {1\over2} \int_\Omega t^{\mu\nu} \delta(g_{\mu\nu}) \sqrt{g} \; d^4x
- {1\over2} \int_{\partial\Omega} \Theta(\delta g) \sqrt{{}_3 g} \; d^3x.
\end{equation}
The surface term, $\Theta$ depends in a linear fashion on $\delta
g$ and its first $n-1$ derivatives, where $n$ denotes the highest
order of the metric derivatives appearing in ${\cal L}$.  In general,
there is no particular reason to expect $\Theta$ to vanish unless
$\delta g$ and its first $n-1$ normal derivatives vanish on the
boundary.  The Einstein-Hilbert Lagrangian is special in this
regard, as is the Lovelock Lagrangian~\cite{Myers}.

\subsection{Lemma: volume term versus surface term}

\subsubsection{Static geometries}

For clarity, I shall first discuss the case of a static, possibly
aspheric, geometry.  Consider the object $\int_\Omega\varrho \sqrt{g}
d^4x$. For the time being, let $\Omega$ denote a four-volume that
is bounded by hypersurfaces of constant lapse $N$. Let $\partial\Omega$
denote its three-boundary, whose normal is by construction orthogonal
to the Killing flow.  Note that by definition $\varrho = t^{\mu\nu}
V_\mu V_\nu = t^{\mu\nu}\; \nabla_\mu t \; K_\nu$. Thus by considering
$\delta(g_{\mu\nu}) \equiv \epsilon V_\mu V_\nu$, one has
\begin{eqnarray}
\int_\Omega\varrho \sqrt{g} d^4x
&=& \int_\Omega  t^{\mu\nu} \; V_\mu V_\nu \; \sqrt{g} d^4x \\
&=& {d\over d\epsilon}\left[ 2 I_m(g+\epsilon V V)
    +  \int_{\partial\Omega}
       \Theta(\delta g= \epsilon V\otimes V) \sqrt{{}_3 g} \; d^3x \right]_.
\end{eqnarray}
The derivative in the above equation is to be evaluated at
$\epsilon=0$.  Introduce the notation $g_\epsilon \equiv g + \epsilon
V V$. Differentiation yields
\begin{eqnarray}
{d\over d\epsilon}\left[ I_m(g+\epsilon V V)\right]
&=& {d\over d\epsilon}
   \int_\Omega  {\cal L}(g_\epsilon) \sqrt{g_\epsilon} \; d^4x
   \nonumber\\
&=& {1\over2} \int_\Omega  {\cal L} \sqrt{g} \; d^4x
    +  \int_\Omega  {d{\cal L}\over d\epsilon} \sqrt{g} \; d^4x.
\end{eqnarray}
Again, everything in the above equation is evaluated at $\epsilon=0$.
Now note that the substitution $g \mapsto g_\epsilon \equiv g +
\epsilon V \otimes V$ merely corresponds to a coordinate change,
a rescaling of the time direction by an amount $\sqrt{1+\epsilon}$.
[One might also profitably think of this as a time dilation.] To
be more explicit
\begin{equation}
\delta g_{\mu\nu} =
\epsilon V_\mu V_\nu =
\epsilon \nabla_{(\mu}t \; K_{\nu)} =
\nabla_{(\mu} [\epsilon t \; K]_{\nu)}.
\end{equation}
Under a coordinate change $x^\mu \mapsto x^\mu + \xi^\mu(x)$, any
arbitrary scalar transforms as $\delta{\cal L} = \xi^\mu(x)
\partial_\mu {\cal L}$. Because the particular coordinate change
under consideration is parallel to the Killing vector, the value
of the Lagrangian is unaltered.  That is: ${d{\cal L}/ d\epsilon}
= 0$. Introduce the notation $f_{\mu\nu} = V_\mu V_\nu = \nabla_{(\mu}t
\; K_{\nu)}$.  Consequently, for any four-volume $\Omega$, bounded
by constant lapse hypersurfaces,  one has:
\begin{equation}
\int_\Omega \{ \varrho - {\cal L}  \} \; \sqrt{g} \; d^4x =
\int_{\partial\Omega} \Theta(\delta g = f)
        \sqrt{{}_3 g} \; d^3x.
\end{equation}

\subsubsection{Stationary geometries}

One must now repeat a minor variant of the above analysis, with
additional technical complications to take care of the black hole's
rotation. Recall that by the assumed internal equilibrium of the
distribution one can show~\cite{Visser93} that the stress energy
tensor has as one of its eigenvectors the four-velocity of the
ROTOR system of co-rotating observers, $V_\Omega$, with the associated
eigenvalue being the energy density, $\varrho$. Indeed
\begin{equation}
 t^\mu{}_\nu (V_\Omega)^\nu = \varrho (V_\Omega)^\mu.
\end{equation}
Now it is certainly true, but not useful, to observe that in the
stationary case $\varrho = t^{\mu\nu} (V_\Omega)_\mu (V_\Omega)_\nu$.
The reason that it is not useful is that explicit computation shows
that it is not possible to interpret $\delta g = (V_\Omega) \otimes
(V_\Omega)$ in terms of the effects of a coordinate transformation.

This is, fortunately, only a technical difficulty and not a
fundamental problem. Introduce the notation $V^\perp$ to denote
some arbitrary four-vector that is constrained only by the fact
that it is assumed to be perpendicular to $V_\Omega$.  That is
$V^\perp \cdot V_\Omega \equiv 0$. Then, because $V_\Omega$ is an
eigenvector of the stress-energy tensor, for any such $V^\perp$
one has $\varrho = t^{\mu\nu} (V_\Omega)_\mu [(V_\Omega)_\nu +
(V^\perp)_\nu]$. The trick is to pick $V^\perp$ in some appropriate
manner. Without further ado, consider
\begin{equation}
V^\perp = ||K_\Omega|| dt - V_\Omega.
\end{equation}
Note that $ dt \cdot V_\Omega = dt \cdot K_\Omega/ ||K_\Omega|| =
1/||K_\Omega||$, so that the perpendicularity requirement is indeed
satisfied.  Furthermore, by explicit construction, $V_\Omega \otimes
[V_\Omega + V^\perp ] = K_\Omega \otimes dt$. Consequently
\begin{equation}
\varrho = t^{\mu\nu} (K_\Omega)_\mu (dt)_\nu .
\end{equation}
Now repeat the analysis used for the static case, this time
considering
\begin{equation}
\delta g_{\mu\nu} =
\epsilon \nabla_{(\mu}t \; [K_\Omega]_{\nu)} =
\nabla_{(\mu} \{\epsilon t \; [K_\Omega]\}_{\nu)}.
\end{equation}
This is nothing more than the effect of the coordinate change $x^\mu
\mapsto x^\mu + \epsilon t [K_\Omega]^\mu$. Consequently the logic
of the preceding case continues to hold, and the lemma is not
disturbed by the black hole's rotation. For this stationary case
introduce the notation $f_{\mu\nu} = \nabla_{(\mu}t \; [K_\Omega]_{\nu)}$.
Again, for any four-volume $\Omega$, bounded by constant lapse
hypersurfaces, one has:
\begin{equation}
\int_\Omega \{ \varrho - {\cal L}  \} \; \sqrt{g} \; d^4x =
\int_{\partial\Omega} \Theta(\delta g = f)
        \sqrt{{}_3 g} \; d^3x.
\end{equation}

\subsection{Entropy}

The preceding lemma essentially solves the problem. Apply the lemma
to the volume term in the entropy formula. As one pushes $\Omega$
outward to cover the whole Euclidean four-manifold two potential
sources of surface term should be considered: surface terms arising
at spatial infinity, and surface terms arising at the horizon. The
surface terms arising at spatial infinity should be quietly discarded
by the assumed asymptotic flatness of spacetime.  The only remaining
piece is the boundary term at the horizon ({\sl
cf}~\cite{Kallosh-Ortin-Peet}). A suitably careful definition of
the entropy is in terms of the limit
\begin{equation}
S = {k A_H\over4 \ell_P^2}
  + {k\over\hbar}
     \lim_H \int_{\partial\Omega} \Theta(f) \sqrt{ {}_3 g} d^3x.
\end{equation}
The limiting procedure, ${\partial\Omega\to H\times[0,\hbar\beta]}$,
may be handled in terms of surfaces of constant lapse function $N$.
[One could just as easily work with hypersurfaces of constant
$N_\Omega = ||K_\Omega||$. Nothing is gained or lost by such a
choice.] Then $\sqrt{ {}_3 g} \; d^3x \mapsto N  \sqrt{ {}_2 g} \;
d^2x \; dt$.  The $t$ integration runs over the range $[0,\hbar\beta]$.
Thus, as one approaches the event horizon
\begin{equation}
S = {k A_H\over4 \ell_P^2}
  +  k \int_H  \beta\lim_H \left[N \Theta(f)\right]
     \; \sqrt{ {}_2 g} \;  d^2x.
\end{equation}
This can be interpreted in terms of a surface entropy density
defined on the event horizon:
\begin{equation}
{\cal S} = k \beta\lim_H \left[N \Theta(f)\right]_.
\end{equation}
Whence
\begin{equation}
S = {k A_H\over4 \ell_P^2}
  + \int_H {\cal S} \; \sqrt{ {}_2 g} \; d^2x.
\end{equation}

An interesting non-zero result is obtained only if $\Theta(f)$ blows
up as $N\to 0$. To see why and when this occurs requires a deeper
understanding of the surface term.  This is as far as I have
currently been able to push the program in the general case. Further
advances seem to require the choice of some specific class of
Lagrangian as template.

Finally, let us reinstate the (volume) entropy density term associated
with the statistical and quantum fluctuations occurring outside the
black hole event horizon. Then
\begin{equation}
S = {k A_H\over4 \ell_P^2}
  +  \int_H  {\cal S} \; \sqrt{ {}_2 g}\; d^2x
  +  \int_\Sigma s \; \sqrt{{}_3 g} \; d^3x.
\end{equation}
Insofar as the quantum fluctuations can be described by some
effective Lagrangian ${\cal L}_{eff}$, they may be extracted from the
volume density $s$, and pushed into the surface density term ${\cal
S}$. This trade-off between volume and surface effects parallels
the trade-off between integrating out fast modes (and describing
them by an effective Lagrangian), and keeping the slow modes
available for explicit computation.

\section{SPECIFIC EXAMPLES}

\subsection{${\cal L} = {\cal L}(Riemann)$}

Consider now the case where one takes ${\cal L}$ to be some arbitrary
function of the Riemann tensor, (though not of its derivatives).
Interest in this class of Lagrangians is justified on the grounds
that {\sl any} quantum theory of gravity will induce terms of this
type in the low-energy effective theory. Many of the examples
considered in the literature are special cases of this reasonably
large class. By the preceding general analysis, evaluation of the
entropy is equivalent to the determination of the value of the
surface term $\Theta$ at the horizon. This surface term
is best evaluated by indirection.


\noindent Define the object:
\begin{equation}
J^{\mu\nu}{}_{\lambda\rho}
\equiv {\partial {\cal L} \over
        \partial R_{\mu\nu}{}^{\lambda\rho} }.
\end{equation}
So that in particular
\begin{equation}
\delta({\cal L}) = J^{\mu\nu}{}_{\lambda\rho} \;
                    \delta(R_{\mu\nu}{}^{\lambda\rho})_.
\end{equation}
Without loss of generality one may take $J^{\mu\nu}{}_{\lambda\rho}$
to inherit the symmetry structure of the Riemann tensor itself.
Specifically $J^{\mu\nu}{}_{\lambda\rho} = J^{[\mu\nu]}{}_{[\lambda\rho]}
=  J_{\lambda\rho}{}^{\mu\nu}$. Consider a general variation of
the metric. Define $\delta g^\mu{}_\nu = g^{\mu\sigma} \delta
g_{\sigma\nu}$. One has
\begin{equation}
\delta(R_{\mu\nu}{}^{\lambda\rho}) =
-2 \nabla_{[\mu} \nabla^{[\lambda} (\delta g)_{\nu]}{}^{\rho]}
+R_{\mu\nu}{}^{\sigma[\lambda} \delta g_\sigma{}^{\rho]}{}.
\end{equation}
This allows us to write
\begin{equation}
\int_\Omega \delta({\cal L}) \sqrt{g} \; d^4x =
\int_\Omega
       J^{\mu\nu}{}_{\lambda\rho}
       \left\{
       -2 \nabla_{\mu} \nabla^{\lambda} (\delta g)_{\nu}{}^{\rho}
       +R_{\mu\nu}{}^{\sigma\lambda} (\delta g)_\sigma{}^{\rho}
       \right\}
\sqrt{g} \; d^4x.
\end{equation}
Here one has been able to drop the explicit antisymmetrization in
view of the symmetry properties of $J^{\mu\nu}{}_{\lambda\rho}$
itself.

{}From the above, one reads off
\begin{eqnarray}
t^{\mu\nu} &=&
-2 \nabla_{\alpha} \nabla_{\beta} J^{\alpha\mu\beta\nu}
  -2 \nabla_{\alpha} \nabla_{\beta} J^{\alpha\nu\beta\mu}
\nonumber \\
&&\quad - J^{\alpha\beta\gamma\mu} R_{\alpha\beta\gamma}{}^\nu
  - J^{\alpha\beta\gamma\nu} R_{\alpha\beta\gamma}{}^\mu
  + {\cal L} g^{\mu\nu} {}.
\end{eqnarray}
Recall the notation
\begin{equation}
f_{\mu\nu} \equiv [K_\Omega]_{(\mu} \nabla_{\nu)}  t.
\end{equation}
For the static case interpret $\Omega_H = 0$, and discard the axial
symmetry. Thus this definition is seen to make sense for both the
static [aspheric] and stationary axisymmetric cases. In either case
we have by construction $\varrho = t^{\mu\nu} f_{\mu\nu}$. Construct
the integral
\begin{eqnarray}
{\cal X}
&\equiv& \int_\Omega     \{\varrho - {\cal L} \} \sqrt{g} d^4x
         \nonumber \\
&=& \int_\Omega \{t^{\mu\nu} \, f_{\mu\nu} - {\cal L}\}\; \sqrt{g} \; d^4x.
\end{eqnarray}
Then
\begin{equation}
{\cal X} =
\int_\Omega
    \{ -4 \nabla_{\alpha} \nabla_{\beta} J^{\alpha\mu\beta\nu}
       - 2 J^{\alpha\beta\gamma\mu} R_{\alpha\beta\gamma}{}^\nu
    \} f_{\mu\nu} \; \sqrt{g} d^4x.
\end{equation}
Integrate by parts once
\begin{eqnarray}
{\cal X}
&=&\int_\Omega
       \left\{
         +4 \nabla_{\beta} J^{\alpha\mu\beta\nu}
         \nabla_{\alpha} ( f_{\mu\nu} )
       \right\} \nonumber\\
     &&\qquad + \left\{
          -2 J^{\alpha\beta\gamma\mu} R_{\alpha\beta\gamma}{}^\nu
            f_{\mu\nu}
       \right\}
\sqrt{g} \; d^4x  \nonumber \\
 &-&4 \int_{\partial\Omega}
   \left\{
      n_\alpha (\nabla_{\beta} J^{\alpha\mu\beta\nu})  f_{\mu\nu}
    \right\}
   \sqrt{{}_3 g} \; d^3x.
\end{eqnarray}
Integrate by parts a second time
\begin{eqnarray}
{\cal X}
&=&\int_\Omega
       \left\{
         -4  J^{\alpha\mu\beta\nu}
         \nabla_{\beta} \nabla_{\alpha} ( f_{\mu\nu} )
       \right\} \nonumber\\
     &&\quad+ \left\{
          -2 J^{\alpha\beta\gamma\mu} R_{\alpha\beta\gamma}{}^\nu
            f_{\mu\nu}
       \right\}
\sqrt{g} \; d^4x \nonumber \\
 &-&4 \int_{\partial\Omega}
   \left\{
      n_\alpha (\nabla_{\beta} J^{\alpha\mu\beta\nu}) f_{\mu\nu}
      - n_\beta J^{\alpha\mu\beta\nu}
        (\nabla_{\alpha}  f_{\mu\nu} )
  \right\}
   \sqrt{{}_3 g} \; d^3x.
\end{eqnarray}
Rearrange
\begin{eqnarray}
{\cal X}
&=&
-2 \int_\Omega  J^{\mu\nu\lambda\rho}
       \left\{
         2 \nabla_{\mu} \nabla_{\lambda} ( f_{\nu\rho} )
         + R_{\mu\nu\lambda}{}^\sigma   f_{\rho\sigma}
       \right\}
\sqrt{g} \; d^4x \nonumber\\
&-&4 \int_{\partial\Omega}
   \left\{
      n_\alpha (\nabla_{\beta} J^{\alpha\mu\beta\nu}) ( f_{\mu\nu} )
      -  J^{\alpha\mu\beta\nu}
        (\nabla_{\alpha}  f_{\mu\nu} ) n_\beta
  \right\}
   \sqrt{{}_3 g} \; d^3x
\end{eqnarray}
The volume integral above vanishes identically. To see this, note
that after appropriate explicit antisymmetrization the volume term
is just
\begin{equation}
\int_\Omega  J^{\mu\nu\lambda\rho} \;
              \delta(R_{\mu\nu\lambda\rho}) \; \sqrt{g} \; d^4x.
\end{equation}
Where $\delta(R_{\mu\nu\lambda\rho})$ is just that due to taking
$\delta(g_{\mu\nu}) = f_{\mu\nu}$. But, as we have already seen
$f_{\mu\nu} = \nabla_{(\mu} [t K_\Omega]_{\nu)}$, which corresponds
to just the effect of a coordinate transformation.

The evaluation of the surface terms proceeds as follows.  First
note that the surface term at spatial infinity is automatically
suppressed by the assumption of asymptotic flatness. Second,
near the horizon $\sqrt{{}_3 g} \; d^3x \to N \sqrt{{}_2 g} \; d^2x
dt$. Since the Riemann tensor and its derivatives are well behaved
at the horizon, as are the limits of $n_\alpha$ and $f_{\mu\nu}$,
it is easy to see that the first surface term vanishes, being
suppressed by the factor of $N$ in the metric determinant.

The only remaining term is
\begin{equation}
{\cal X}
=
4 \lim_H \int_{\partial\Omega}
   \left\{
         J^{\alpha\mu\beta\nu}
         \nabla_{\alpha}( f_{\mu\nu} ) n_\beta
  \right\}
   \sqrt{{}_3 g} \; d^3x.
\end{equation}
This formula is, of course, nothing more nor less than the special
case explicit evaluation of the surface term $\Theta(f)$ previously
encountered in the general argument.

At this stage it proves useful to treat the stationary and static
cases separately.

\noindent\underline{Static geometry:}

For a static geometry $f_{\mu\nu} = V_\mu V_\nu$. The gradient term
includes pieces such as
\begin{equation}
\nabla_{\alpha} (V_\mu V_\nu)
= (\nabla_{\alpha} V_\mu) V_\nu + V_\mu (\nabla_\alpha V\nu )
= -||a|| \left( V_\alpha n_\mu V_\nu + V_\mu  V_\alpha n_\nu \right).
\end{equation}
Now, take the limit as one approaches the horizon. The only surviving
term in the surface integral comes from the cancellation between
the $N$ arising from the metric determinant and $||a||$. Note that
\begin{equation}
\lim_H \int_0^{\hbar\beta} dt N ||a|| = \lim_H \hbar\beta N ||a|| = 2\pi.
\end{equation}
This yields
\begin{equation}
\int_\Omega \{\varrho - {\cal L} \} \sqrt{g} \; d^4x =
8\pi \int_H
  \left\{
        J^{\alpha\mu\beta\nu} V_\alpha V_\beta n_\mu n_\nu
  \right\}
   \sqrt{{}_2 g} \; d^2x.
\end{equation}

\noindent\underline{Stationary geometry:}

The limiting procedure is now a little more delicate, and requires
some tedious technical steps. Recall that $f_{\mu\nu} = [K_\Omega]_{(\mu}
[dt]_{\nu)}$.  The gradient term includes pieces such as
\begin{equation}
\nabla_{\alpha} ([K_\Omega]_\mu [dt]_\nu)
=
(\nabla_{\alpha} [K_\Omega]_\mu) [dt]_\nu
+ [K_\Omega]_\mu (\nabla_\alpha \nabla_\nu t).
\end{equation}
Because the Killing vector $K_\Omega$ is not hypersurface orthogonal,
in general, the best we can say is that the  covariant derivative
of the Killing vector satisfies
\begin{equation}
\nabla_\mu [K_\Omega]_\nu
= 2 (a_\Omega)_{[\mu} (K_\Omega)_{\nu]} + \pi_{\mu\nu}.
\end{equation}
Here $\pi_{\mu\nu}$ is an antisymmetric tensor orthogonal to
$K_\Omega$.  On the other hand, it is known that $K_\Omega$ is
hypersurface orthogonal on the event horizon, so that $\pi_{\mu\nu}$
vanishes in that limit. Consequently
\begin{equation}
\nabla_\mu [K_\Omega]_\nu
= -2 N_\Omega ||a_\Omega|| n_{[\mu} (V_\Omega)_{\nu]} + O(N_\Omega).
\end{equation}
Now consider
\begin{eqnarray}
(\nabla_\mu \nabla_\nu t)
 &=& \nabla_\mu [(K_\nu + \omega\tilde K_\nu)/N^2 ]\nonumber\\
 &=& -2 N^{-3} \nabla_{(\mu} N \;[K + \omega\tilde K]_{\nu)}
     +  N^{-2} \nabla_{(\mu} \omega \;\tilde K_{\nu)} \nonumber\\
 &=& -2 N^{-1} [a_\omega]_{(\mu}  \;[V_\omega]_{\nu)}
     +  N^{-2} \nabla_{(\mu} \omega \;\tilde K_{\nu)} \nonumber\\
&=& +2 {||a_\omega|| \over N} \; [n_\omega]_{(\mu}  \;[V_\omega]_{\nu)}
     +  N^{-2} \nabla_{(\mu} \omega \;\tilde K_{\nu)}.
\end{eqnarray}
Now because $\omega = \Omega_H + O(N^2)$ one has $\nabla\omega =
O(N\nabla N) = O(N^2 a)$. So finally
\begin{equation}
(\nabla_\mu \nabla_\nu t)
= +2 {||a_\omega|| \over N} \;
       \left\{ [n_\omega]_{(\mu}  \;[V_\omega]_{\nu)} + O(N) \right\}_.
\end{equation}
Now, take the limit as one approaches the horizon. The ZAMOS,
$V_\omega$, and the ROTORS, $V_\Omega$, approach the same limit.
Ditto for the relevant normals.  Again, the only surviving term in
the surface integral comes from the cancellation between the $N$
arising from the metric determinant $||a_\omega||$, and $||a_\Omega||$.
Various subdominant pieces vanish in the limit.
As before, this yields
\begin{equation}
\int_\Omega \{\varrho - {\cal L} \} \sqrt{g} \; d^4x =
8\pi \int_H
  \left\{
        J^{\alpha\mu\beta\nu} V_\alpha V_\beta n_\mu n_\nu
  \right\}
   \sqrt{{}_2 g} \; d^2x.
\end{equation}

Returning to the general case (static or stationary), and backtracking
to the general entropy formula, one now obtains
\begin{equation}
S = {k A_H\over4\ell_P^2}
  + 8\pi {k\over\hbar} \int_H \;
         J^{\mu\nu\lambda\rho} \;
         V_\mu n_\nu V_\lambda n_\rho
	\; \sqrt{ {}_2 g} \; d^2x .
\end{equation}
A further refinement is to define $g^\perp_{\mu\nu} \equiv V_\mu
V_\nu + n_\mu n_\nu$. This is essentially the metric in the two
directions perpendicular to the event horizon, in terms of which
the symmetries of $J^{\mu\nu\lambda\rho}$ imply
\begin{equation}
S = {k A_H\over4\ell_P^2}
  + 4\pi {k\over\hbar} \int_H \;
         J^{\mu\nu\lambda\rho} \;
	 g^\perp_{\mu\lambda}g^\perp_{\nu\rho} \;
	\sqrt{ {}_2 g} \; d^2x .
\end{equation}
This is our final form for the entropy. Note that it exhibits all
of the properties expected from the general analysis.

\subsection{${\cal L} = {\cal L}(Ricci)$}

A further specialization of the above, is to consider the case
where the Lagrangian is an arbitrary function of the Ricci tensor,
rather than the full Riemann tensor. The analysis is straightforward.

\noindent Define
\begin{equation}
{\tilde J}^{\mu\nu}
\equiv {\partial {\cal L} \over
        \partial R_{\mu\nu}}.
\end{equation}
Then
\begin{equation}
J^{\mu\nu}{}_{\lambda\rho}
\equiv {\partial {\cal L} \over \partial R_{\mu\nu}{}^{\lambda\rho} }
=  {\tilde J}^{[\mu}{}_{[\lambda} g^{\nu]}{}_{\rho]} =
{1\over4}\left[ {\tilde J}^\mu{}_\lambda g^\nu{}_\rho
               -{\tilde J}^\mu{}_\rho g^\nu{}_\lambda
	       +{\tilde J}^\nu{}_\rho g^\mu{}_\lambda
	       -{\tilde J}^\nu{}_\lambda g^\mu{}_\rho \right]_.
\end{equation}
Insert into the previous formula, one extracts
\begin{equation}
S = {k A_H\over4\ell_P^2}
  + 2\pi {k\over\hbar} \int_H \;
    {\partial {\cal L} \over \partial R_{\mu\nu}} \; g^\perp_{\mu\nu} \;
    \sqrt{ {}_2 g} d^2x.
\end{equation}
This formula is instructively similar to that obtained by Jacobson, Kang,
and Myers~\cite{JM:talk,JM:93b}. Using field redefinition techniques
under conformal rescalings they were limited to the case ${\cal L}
= R_{\mu\nu} R^{\mu\nu} + {\cal L}_{matter}$. (The extra matter
was required to enforce a nontrivial solution to the fields equations,
it was assumed that the extra matter was sufficiently well behaved
not to contribute to the entropy in its own right.) Under these
assumptions, Jacobson, Kang, and Myers showed that
\begin{equation}
S = {k A_H\over4\ell_P^2}
  + 4\pi {k\over\hbar} \int_H \;
    R^{\mu\nu} \; g^\perp_{\mu\nu} \; \sqrt{ {}_2 g} d^2x.
\end{equation}

\subsection{${\cal L} = {\cal L}(Tr[Ricci])$}

A completely analogous analysis can be applied in the case that
the Lagrangian  is an arbitrary  function of the scalar  Ricci
curvature.

\noindent Consider
\begin{equation}
J^{\mu\nu}{}_{\lambda\rho}
\equiv {\partial {\cal L} \over
        \partial R_{\mu\nu}{}^{\lambda\rho} }
=  {\partial {\cal L} \over \partial R} \;
   g^{[\mu}{}_{[\lambda} g^{\nu]}{}_{\rho]}.
\end{equation}
Insert into the general result, one obtains
\begin{equation}
S = {k A_H\over4\ell_P^2}
  + 4\pi {k\over\hbar} \int_H \; {\partial {\cal L} \over \partial R}
     \; \sqrt{ {}_2 g} d^2x.
\end{equation}
This is exactly the result enunciated by Jacobson, Kang, and
Myers~\cite{JM:talk,JM:93b}.

A simple consistency check is to lump the Einstein-Hilbert action
in with ${\cal L}$. Taking ${\cal  L} = {1\over16\pi G} R =
{\hbar\over16\pi\ell_P^2} R$ reproduces the ordinary area term.

\subsection{Lovelock gravity}

As a final example, I shall discuss Lovelock gravity. While the
analysis presented so far has, for definiteness, been presented in
four dimensions there is nothing essentially four-dimensional about
these techniques. In $D$ dimensions the Lovelock Lagrangian is
given by (see {\sl e.g.}~\cite{JM:93a}).
\begin{equation}
{\cal L}={\sum_{m=0}^{[D/2]}} c_m \; {\cal L}_m.
\end{equation}
In this sum, $[D/2]$ indicates the integer part of $D/2$.  The
individual terms  are given by
\begin{equation}
{\cal L}_m(g)
={1\over2^m} \;
\delta_{\lambda_1\rho_1\cdots\lambda_m\rho_m}^{\mu_1\nu_1\cdots \mu_m\nu_m}
\;\;
R_{\mu_1\nu_1}{}^{\lambda_1\rho_1}\cdots R_{\mu_m\nu_m}{}^{\lambda_m\rho_m}{}.
\end{equation}
The $\delta$ symbol is a totally antisymmetric product of $2m$
Kronecker deltas, suitably normalized to take values $0$ and $\pm
1$. It is convenient to define ${\cal L}_0=1$, this term corresponding
to a cosmological constant. Furthermore ${\cal L}_1=R$ is
the Einstein-Hilbert Lagrangian.  In general, ${\cal L}_m$ is the
Euler density for a $2m$-dimensional manifold.  Because of the
antisymmetrization, no derivative appears at higher than second
order in the equations of motion.

For the purposes currently at hand, consider the object
\begin{eqnarray}
[J_m]^{\mu\nu}{}_{\lambda\rho}
&\equiv& {\partial {\cal L}_m \over
        \partial R_{\mu\nu}{}^{\lambda\rho} }
	\nonumber\\
&=& {m\over2^m} \;
\delta_{\lambda_1\rho_1\cdots
        \lambda_{m-1}\rho_{m-1}\lambda\rho}^{\mu_1\nu_1\cdots
	\mu_{m-1}\nu_{m-1}\mu\nu}\;\;
         R_{\mu_1\nu_1}{}^{\lambda_1\rho_1}\cdots
	 R_{\mu_{m-1}\nu_{m-1}}{}^{\lambda_{m-1}\rho_{m-1}}_.
\end{eqnarray}
Applying the general formula, the contractions with $g^\perp$,
together with the total antisymmetrization of the indices, imply
that the only components of Riemann tensor that contribute to the
entropy density are those that are tangential to the $D-2$ dimensional
event horizon. Specifically, we note that
\begin{equation}
\delta_{\lambda_1\rho_1
        \cdots\lambda_{m-1}\rho_{m-1}\lambda\rho}^{\lambda_1\rho_1\cdots
	\mu_{m-1}\nu_{m-1}\mu\nu}\;\;
(g^\perp)^{[\lambda}{}_{[\mu} (g^\perp)^{\rho]}{}_{\nu]}  =
{\tilde\delta}_{\lambda_1\rho_1\cdots
                \lambda_{m-1}\rho_{m-1}}^{\mu_1\nu_1\cdots
		\mu_{m-1}\nu_{m-1}}.
\end{equation}
Here $\tilde \delta$ is the totally antisymmetric product of $2(m-1)$
Kronecker deltas, restricted to the subspace orthogonal to $g^\perp$.

The rest of the derivation now parallels that due to Jacobson and
Myers~\cite{JM:93a}.  The entropy is
\begin{equation}
S = 4\pi {k\over\hbar}
   \int_H {\sum_{m=1}^{[D/2]}} \;
     m \; c_m \; {\cal L}_{m-1}(h) \; \sqrt{h} \; d^{D-2}x.
\end{equation}
In this particular case the entropy is given solely in terms of
the intrinsic geometry ($h$) of the event horizon --- this result is
special to this particular type of Lagrangian and does not
generalize.

\section{DISCUSSION}

The computation of black hole entropies in various model theories
is an issue of great current interest. Based on the work of a
several authors the situation has by now become significantly
clarified. Several points are worth making.

First: The naive area law for black hole entropies is in general false.
\begin{equation}
S \neq {k A_H\over4\ell_P^2}.
\end{equation}
The naive law certainly holds for Einstein-Hilbert gravity coupled
to matter whose kinetic energy is quadratic~\cite{Visser93}. Once
one moves beyond quadratic kinetic energies the naive law fails in
general.

Second: For a general higher derivative Lagrangian the entropy of
a black hole is given by an integral of some suitable density over
(a fixed time spacelike cross-section of) the event horizon
\begin{equation}
S = {k A_H\over4 \ell_P^2}
  +  \int_H  {\cal S} \; \sqrt{ {}_2 g}\; d^2x.
\end{equation}
The entropy surface density is a simple function of the surface
term that connects the stress-energy tensor with the variation of
the action under a variation of the spacetime metric.

Third: In the specific case of a Lagrangian that is solely a
function of the Riemann tensor
\begin{equation}
S = {k A_H\over4\ell_P^2}
  + 4\pi {k\over\hbar} \int_H \;
         {\partial {\cal L}\over\partial R_{\mu\nu\lambda\rho}} \;
	 g^\perp_{\mu\lambda}g^\perp_{\nu\rho} \;
	\sqrt{ {}_2 g} \; d^2x .
\end{equation}
This relatively general formula can be checked against a number of
more specific examples where the entropy is known by other means.
The fact that different types of calculation give the same answer
where they overlap is certainly encouraging.

Fourth: The present paper has obtained its results via extensive
use of Euclidean signature techniques. The underlying physics is
perhaps somewhat obscured by this formalism. It is encouraging to
note that similar results have by now been presented using a number
of different techniques~\cite{JM:talk,JM:93b,Wald:talk,Wald93}.
The overall agreement between these various different techniques is
a further useful consistency check.

\underline{Acknowledgements}

This research was supported by the U.S. Department of Energy.
I wish to thank Ted Jacobson, Robert Myers, and Robert Wald for
stimulating discussions, and for a reading of the manuscript.

\newpage

\end{document}